\documentclass{aastex63}
\pdfoutput=1 
\usepackage{amsmath,amstext}
\usepackage[T1]{fontenc}
\usepackage{apjfonts} 
\usepackage[figure,figure*]{hypcap}

\newcommand{\p}[1]{{\color{cyan}{#1}}}
\renewcommand{\d}[1]{{\color{red}{#1}}}

\newcommand{\angstrom}{\mbox{\normalfont\AA}}
\newcommand{\g}[1]{{\color{cyan}{#1}}}

\shorttitle{Evolution of Plasma Composition in an Eruptive Flux Rope}
\shortauthors{Baker, D. et al.}

\begin{document}

\title{Evolution of Plasma Composition in an Eruptive Flux Rope}
\author{Baker, D.}
\affiliation{University College London, Mullard Space Science Laboratory, Holmbury St. Mary, Dorking, Surrey, RH5 6NT, UK}
\author{Green, L. M.}
\affiliation{University College London, Mullard Space Science Laboratory, Holmbury St. Mary, Dorking, Surrey, RH5 6NT, UK}
\author{Brooks, D. H.}
\affiliation{College of Science, George Mason University, 4400 University Drive, Fairfax, VA 22030, USA}
\author{D\'emoulin, P.}
\affiliation{LESIA, Observatoire de Paris, Universit\'e PSL, CNRS, Sorbonne Universit\'e, Univ. Paris Diderot, Sorbonne Paris Cit\'e, 5 place Jules Janssen, 92195 Meudon, France}
\affiliation{Laboratoire Cogitamus, rue Descartes, 75005 Paris, France}
\author{van Driel-Gesztelyi, L.}
\affiliation{University College London, Mullard Space Science Laboratory, Holmbury St. Mary, Dorking, Surrey, RH5 6NT, UK}
\affiliation{LESIA, Observatoire de Paris, Universit\'e PSL, CNRS, Sorbonne Universit\'e, Univ. Paris Diderot, Sorbonne Paris Cit\'e, 5 place Jules Janssen, 92195 Meudon, France}
\affiliation{Konkoly Observatory, Research Centre for Astronomy and Earth Sciences, Konkoly Thege \'ut 15-17., H-1121, Budapest, Hungary}
\author{Mihailescu, T.}
\affiliation{University College London, Mullard Space Science Laboratory, Holmbury St. Mary, Dorking, Surrey, RH5 6NT, UK}
\author{To, A. S. H.}
\affiliation{University College London, Mullard Space Science Laboratory, Holmbury St. Mary, Dorking, Surrey, RH5 6NT, UK}
\author{Long, D. M.}
\affiliation{University College London, Mullard Space Science Laboratory, Holmbury St. Mary, Dorking, Surrey, RH5 6NT, UK}
\author{Yardley, S. L.}
\affiliation{University College London, Mullard Space Science Laboratory, Holmbury St. Mary, Dorking, Surrey, RH5 6NT, UK}
\author{Janvier, M.}
\affiliation{Institut d'Astrophysique Spatiale, UMR8617, Univ. Paris-Sud-CNRS, Universit\'e Paris-Saclay, Bâtiment 121, 91405 Orsay Cedex, France}
\author{Valori, G.}
\affiliation{Max Planck Institute for Solar System Research, Justus-von-Liebig-Weg 3, 37077 G\"ottingen, Germany}

\begin{abstract}
Magnetic flux ropes are bundles of twisted magnetic field enveloping a central axis. 
They harbor free magnetic energy and can be progenitors of coronal mass ejections (CMEs), but identifying flux ropes on the Sun can be challenging. 
One of the key coronal observables that has been shown to indicate the presence of a flux rope is a peculiar  bright coronal structure called a sigmoid. 
In this work, we show Hinode EUV Imaging Spectrometer (EIS) observations of sigmoidal active region 10977. 
We analyze the coronal plasma composition in the active region and its evolution as the sigmoid (flux rope) forms and erupts as a CME. 
Plasma with photospheric composition was observed in coronal loops close to the main polarity inversion line during episodes of significant flux cancellation, suggestive of the injection of photospheric plasma into these loops driven by photospheric flux cancellation. 
Concurrently, the increasingly sheared core field contained plasma with coronal composition. 
As flux cancellation decreased and the sigmoid/flux rope formed, the plasma evolved to an intermediate composition in between photospheric and typical active region coronal compositions. 
Finally, the flux rope contained predominantly photospheric plasma during and after a failed eruption preceding the CME. 
Hence, plasma composition observations of active region 10977 strongly support models of flux rope formation by photospheric flux cancellation forcing magnetic reconnection first at the photospheric level then at the coronal level.
\end{abstract}


\section{Introduction}\label{s:intro}

Magnetic flux ropes are specific magnetic configurations in the solar atmosphere where helical field lines wrap around a common axial field.
They are fundamentally associated with solar eruptions, particularly coronal mass ejections (CMEs), due to their magnetic free energy content and susceptibility to a loss of equilibrium or instability \citep[see][for a review]{green18}.
Although the magnetic field of flux ropes cannot readily be directly observed in imaging data, sigmoids are a well-known indirect signature that indicates the presence of helical field lines, of around one turn, in a flux rope  configuration \citep[][]{rust96,green11}.
Sigmoids are hot, $S$-shaped (or double $J$-shaped) coronal loops that emit in EUV and soft X-ray, covering a temperature range of log $T_{K}$ = [6.0, 7.2] \citep[e.g.][]{gibson02,tripathi09,james18,mulay21}. 
When observed on the Sun, sigmoids are highly likely to erupt as a CME \citep{rust96,canfield99,canfield07}.

Flux ropes in sigmoidal active regions can form during an active region's emergence and/or decay phase. 
Regardless of the phase though, flux rope formation can be a consequence of photospheric flows that drive reconnection at some height in the atmosphere. 
For example, during an active region's emergence phase, strong orbiting motions of photospheric field bring together sheared loops systems and drive reconnection between them, resulting in flux rope formation in the corona \citep[][]{james20}. 
Similarly, during the decay phase, reconnection in the photosphere, which manifests itself as flux cancellation \citep{martin85}, readily occurs and transforms an active region's sheared arcade into a low-lying flux rope \citep{vanballegooijen89}. 
This process takes place over several days and once the sigmoid forms as a continuous $S$-shape (from double $J$-shaped loops), a CME follows within a period of time measured in hours \citep{green14}. 
As decaying active regions disperse their fragmented flux over an ever larger area, flux cancellation and flux rope formation readily occur along the internal or main polarity inversion line (PIL) of the region \citep[e.g.][]{green11,green18,yardley18}.  
These transformations of the magnetic configuration are realized with magnetic reconnection occurring from  photospheric up to low coronal heights.
Different reconnection heights ultimately influence the specific details of the flux rope, and the plasma it contains, and its likelihood to erupt as a CME. Therefore, due to the very nature of the formation process of flux ropes, plasma composition is a potentially powerful diagnostic to constrain flux rope formation models, with measured elemental composition of sigmoid plasma providing information as to its origin, whether photospheric or coronal.

Plasma composition can be determined by considering coronal emission lines from elements with different first ionization potentials (FIP). 
In general, elements with a FIP $\lesssim$ 10 eV have enhanced abundances compared to those with a FIP $\gtrsim$ 10 eV when the plasma is observed in the corona relative to the photosphere \citep[see the review of][]{laming15}.
The degree of enhancement is highly correlated with the Sun's magnetic activity on all spatial and temporal scales
\citep[e.g.][]{brooks15,baker18,brooks17}. 
Studies of erupting prominences show that their cool, dense plasma has photospheric composition, suggesting that  unfractionated plasma from the photosphere/chromosphere was brought upwards into the prominence body rather than fractionated plasma from the corona condensing \citep[e.g.][]{feldman92,spicer98,ciaravella00}.
\cite{parenti19} confirmed that quiescent prominences also have photospheric composition.

These studies focused on the properties of the filament/prominence material suspended in a flux rope configuration.
To date there are few examples of plasma composition being used either on its own or with other observational evidence to investigate the formation and evolution of flux ropes in active regions (ARs).
\cite{baker13} found unfractionated plasma, i.e. of photospheric composition, along a sigmoid channel in an eruptive active region.
The authors concluded that the observed photospheric composition plasma combined with significant flux cancellation was evidence of a flux rope that had formed via reconnection low down in the solar atmosphere as proposed by \cite{vanballegooijen89}.
Coronal plasma composition was a key observable used to verify that a sigmoidal flux rope formed via reconnection high up in the corona by \citet{james17,james18}.
\cite{fletcher01} established a link between elemental abundances and the type of magnetic topology associated with transition region brightenings within a sigmoidal active region. 
More precisely, they found that the brightenings related to bald patch separatrix and quasi-separatrix layers are associated with plasma composition close to typical photospheric and coronal composition, respectively.  

In the investigation presented here, coordinated \emph{Hinode}/XRT, SOT, and EIS observations are used to show how plasma composition evolves in a sigmoidal active region as a flux rope forms and then subsequently erupts.
In Section \ref{s:evolution}, we provide observations of the photosphere, chromosphere, and corona during the active region's decay phase when the flux rope formed.
An account of the \emph{Hinode}/EIS diagnostics used in this study follows in Section \ref{s:eis}.
The spatially resolved composition ratio and temperature maps are presented in Section \ref{s:maps}. 
We discuss our findings in the context of flux rope formation models based on flux cancellation \cite[e.g.][]{vanballegooijen89,aulanier10} in Section \ref{s:discussion} before concluding in Section \ref{s:conclusion}.

\section{Evolution of AR 10977}\label{s:evolution}

\begin{figure}[bt!]
\epsscale{1.2}
\plotone{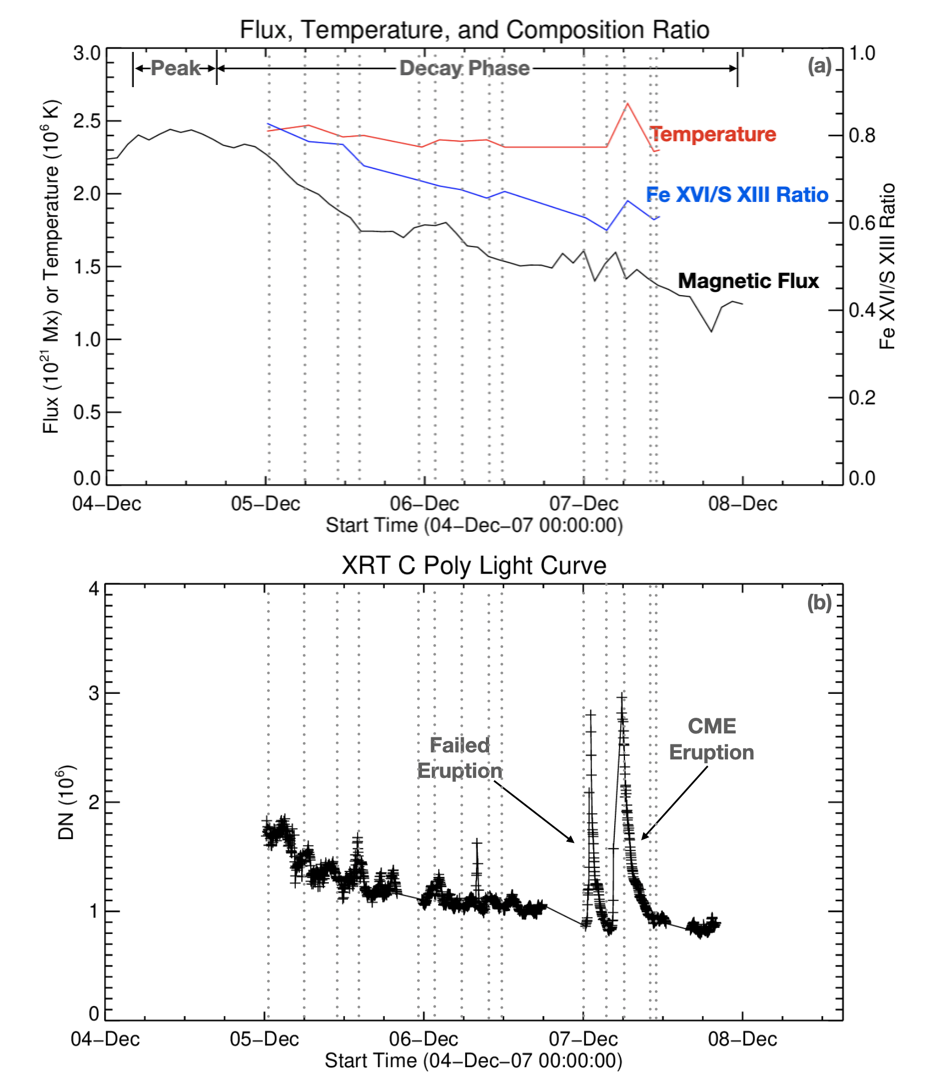}
\caption{Global evolution of AR 10977.
(a) Half of total unsigned flux (black), mean temperature (red), and mean \ion{Fe}{16} 262.98 $\angstrom$ / \ion{S}{13} 256.69 $\angstrom$ composition ratio (blue) vs time. \emph{Hinode}/EIS raster start times are shown with dashed gray lines.
The global evolution of magnetic flux, temperature, and composition ratio are compared in Section \ref{s:Global_Evolution}.
(b) \emph{Hinode}/XRT C Poly light curve for December 5--7. 
\label{fig:global}} 
\end{figure}

\begin{figure*}[hbt!]
\epsscale{1.2}
\plotone{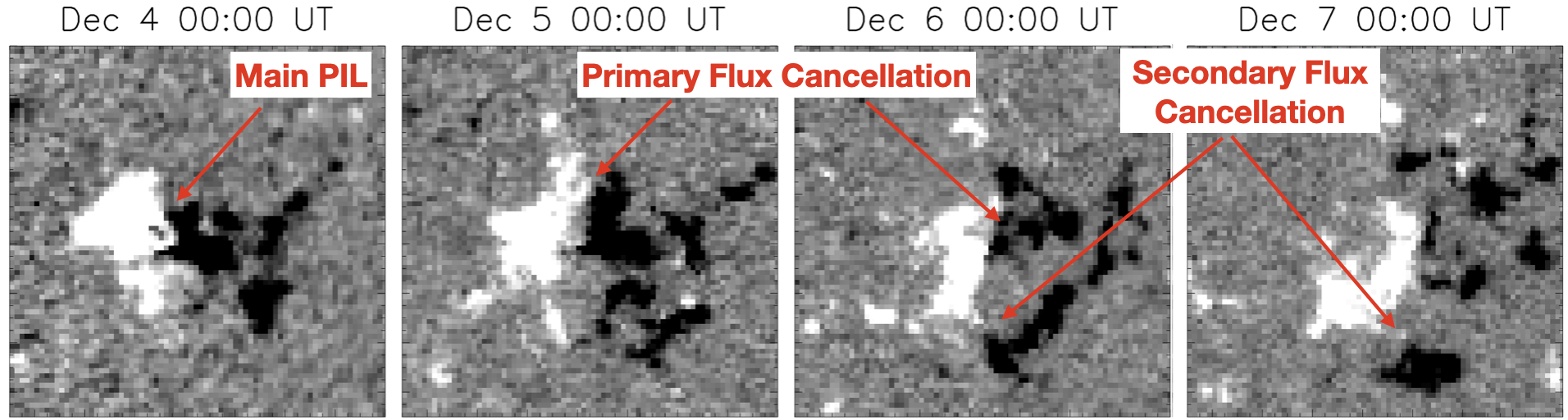}
\caption{SOHO/MDI LOS magnetograms of AR 10977 at 00:00 UT on 2007 December 4--7. 
\label{fig:mdi_series}} 
\end{figure*}
\begin{figure*}[hbt!]
\epsscale{1.2}
\plotone{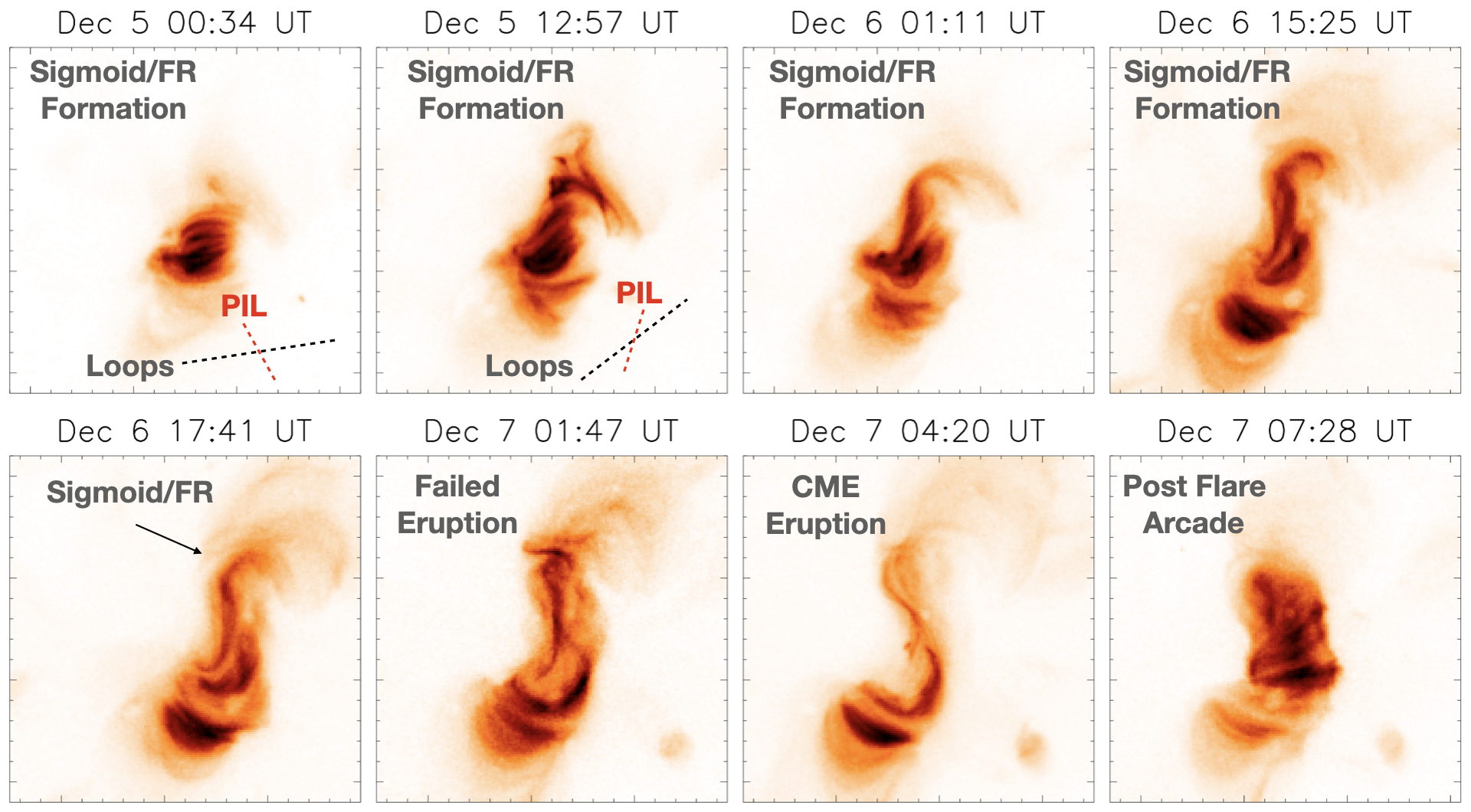}
\caption{\emph{Hinode}/XRT C Poly images of the coronal field evolution during the period of December 5--7. 
A sense of increasing shear in the northern section of active region is indicated by the black and red dashed lines showing the approximate angle between the main PIL and loops crossing it.
A movie of the \emph{Hinode}/XRT images is included as `XRT$\_$movie.mp4'.  The movie covers the time period from 04:39 UT on 2007 December 5 to 12:18 UT on 2007 December 7.  The field of view centers on the sigmoidal active region as it evolves and eventually erupts as a CME. 
\label{fig:xrt_series}} 
\end{figure*}

AR 10977 was a simple bipolar active region (Zurich classification $\beta$) visible on the solar disk from 2007 December 2--12. 
The active region was in its emergence phase for approximately two days before a peak flux value of $\sim$2.4$\times$10$^{21}$ Mx was reached at $\sim$ 12:00 UT on December 4. 
The active region then entered its decay phase, which was characterized by the formation of a sigmoid indicating that a flux rope had been built \citep{green11,savcheva12,gibb14}.
The flux rope erupted early on December 7 in two stages -- a failed eruption followed a few hours later by a CME that was detected in STEREO-B coronagraph data \citep{ma09b}.
A B1.4--class GOES flare and a global wave were associated with the CME \citep[][]{ma09,green11,long11,attrill14}.

Figure \ref{fig:global}(a) shows half of the total unsigned magnetic flux (black curve) at the  end of the emergence phase and during the decay phase, with the times of \emph{Hinode}/EIS observations plotted as vertical gray dashed lines.
During the early decay phase, significant flux cancellation occurred along the northern section of the main PIL for 2.5 days prior to the CME that occurred at 04:20 UT on the 7th \citep{green11}.
At the southern-most end of the main PIL, flux cancellation was also observed in the more fragmented magnetic field but this was minor compared to that of the primary site in the north and cancellation began later there \citep{green11}.  
The locations of the main PIL and the sites of flux cancellation are identified in the SOHO/MDI magnetograms of Figure \ref{fig:mdi_series}. 

Figure \ref{fig:global}(b) shows a light curve of the soft X-ray emission from the entire active region.
The soft X-ray emission associated with the failed eruption and the CME peaked at 01:08 UT and 05:45 UT, respectively, on 2007 December 7.
The evolution of the coronal loops during the region's decay phase can be seen in the \emph{Hinode}/XRT images in Figure \ref{fig:xrt_series}.
The image series shows that the coronal field evolves in three key stages during the decay phase:  flux rope formation,  failed/CME eruptions, and post-CME eruption, briefly described below and discussed in more detail in Sections \ref{stage1}, \ref{stage2}, and \ref{stage3}, respectively.

Early on December 5, before the start of significant flux cancellation in the north, the arcade loops are aligned approximately orthogonal to the main PIL i.e. are potential (see image at 00:34 UT on December 5).
The arcade field is more sheared 12 hours later as flux cancellation is accelerating (see image at 12:57 UT).
The approximate shear angles are shown by the crossing of the dashed red/black lines representing the main PIL/loops axes in both images.
By 15:51 UT on December 6, the active region loops have formed a continuous forward S-shaped sigmoid \citep{green11}.
The sigmoid/flux rope expands and rises during the  failed eurption and CME (see images at 01:47 and 04:20 UT on December 7, respectively). 
Highly sheared post-eruption loops are present in the northern region immediately following the CME from 05:00 to 12:00 UT.
The sigmoid was destroyed during the CME \citep{green11} but reformed after \emph{Hinode}/EIS composition observations ended at 11:26 UT.

\begin{figure}[hbt!]
\epsscale{0.70}
\plotone{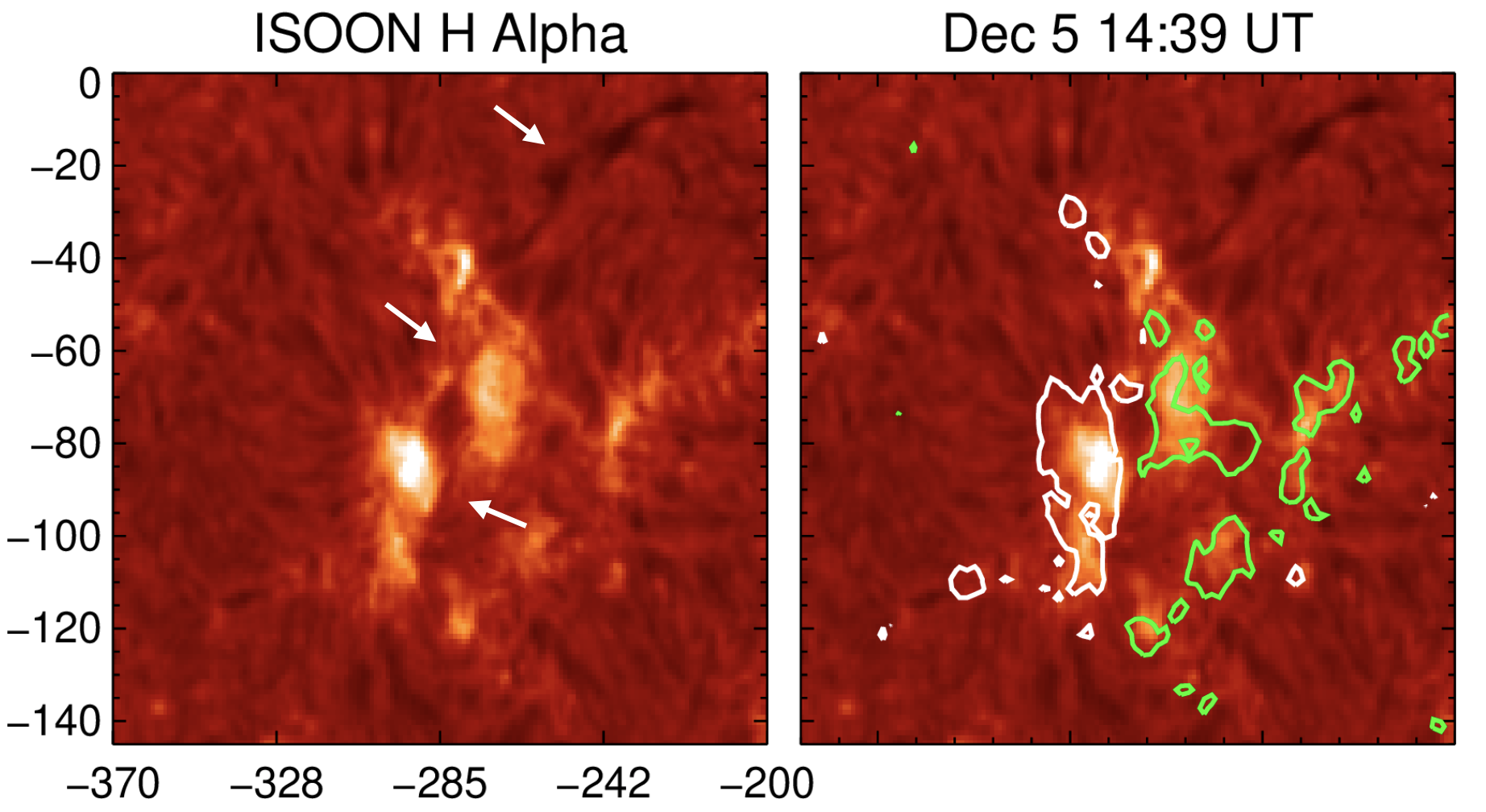}
\plotone{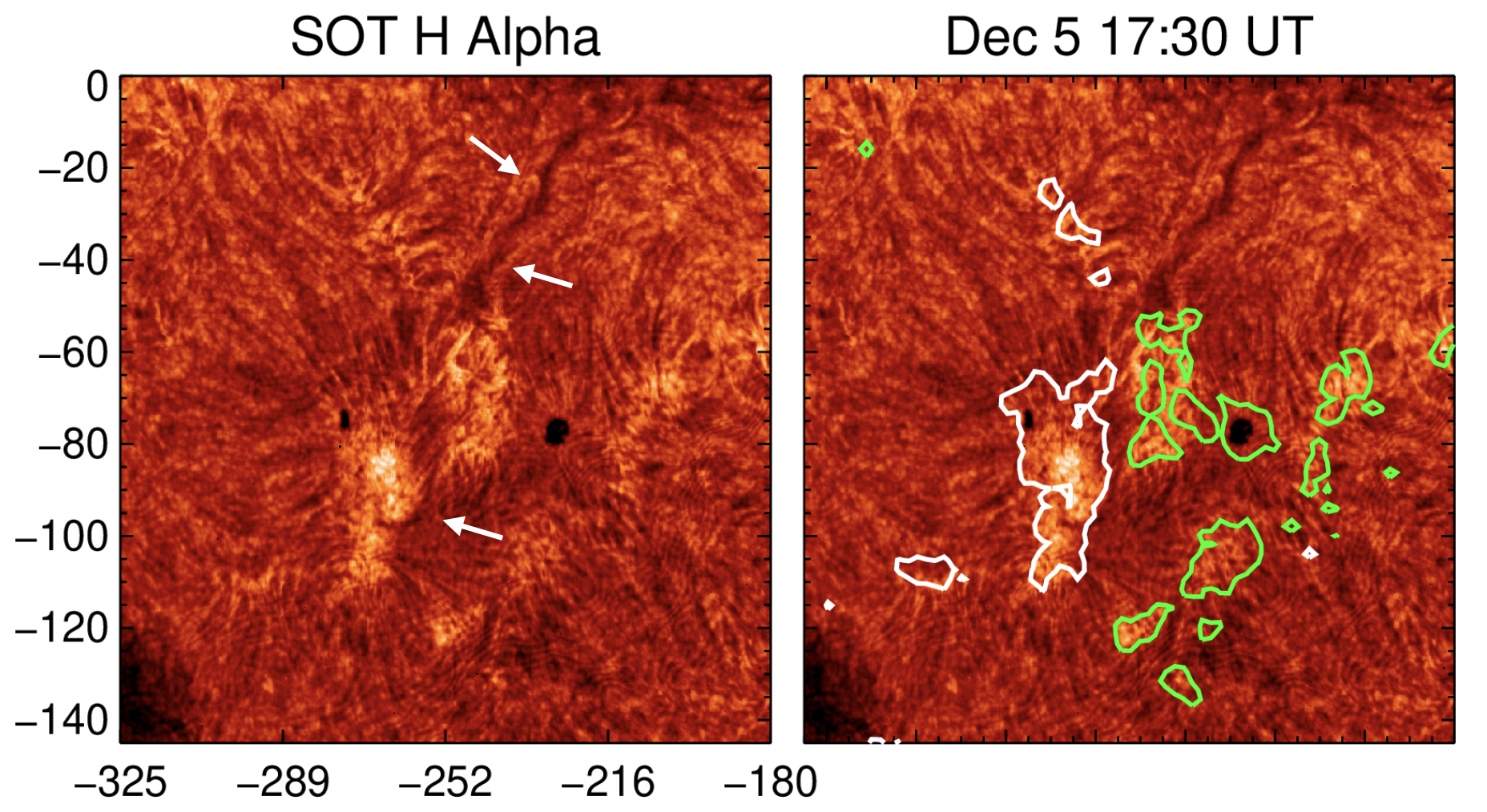}
\plotone{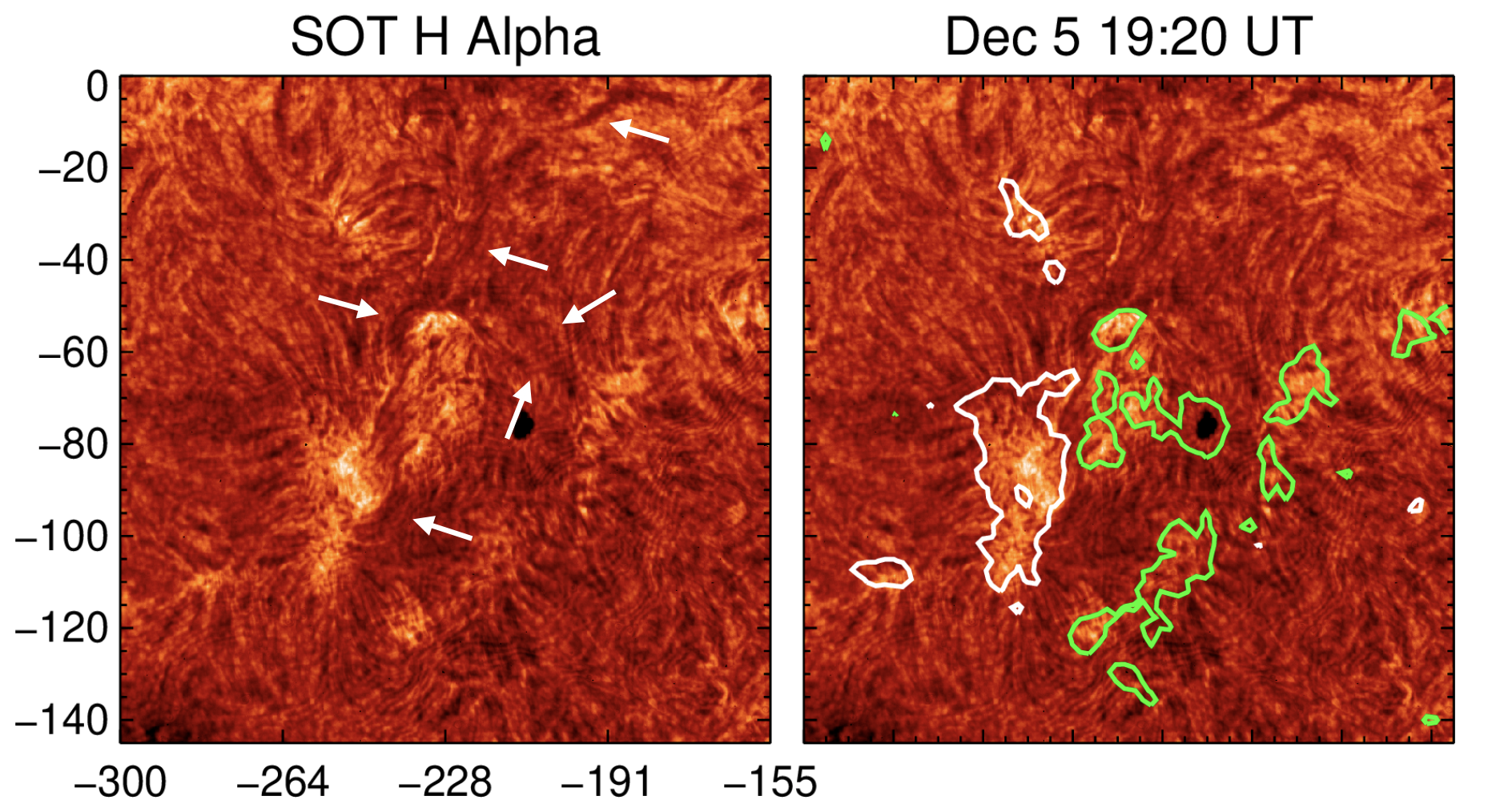}
\plotone{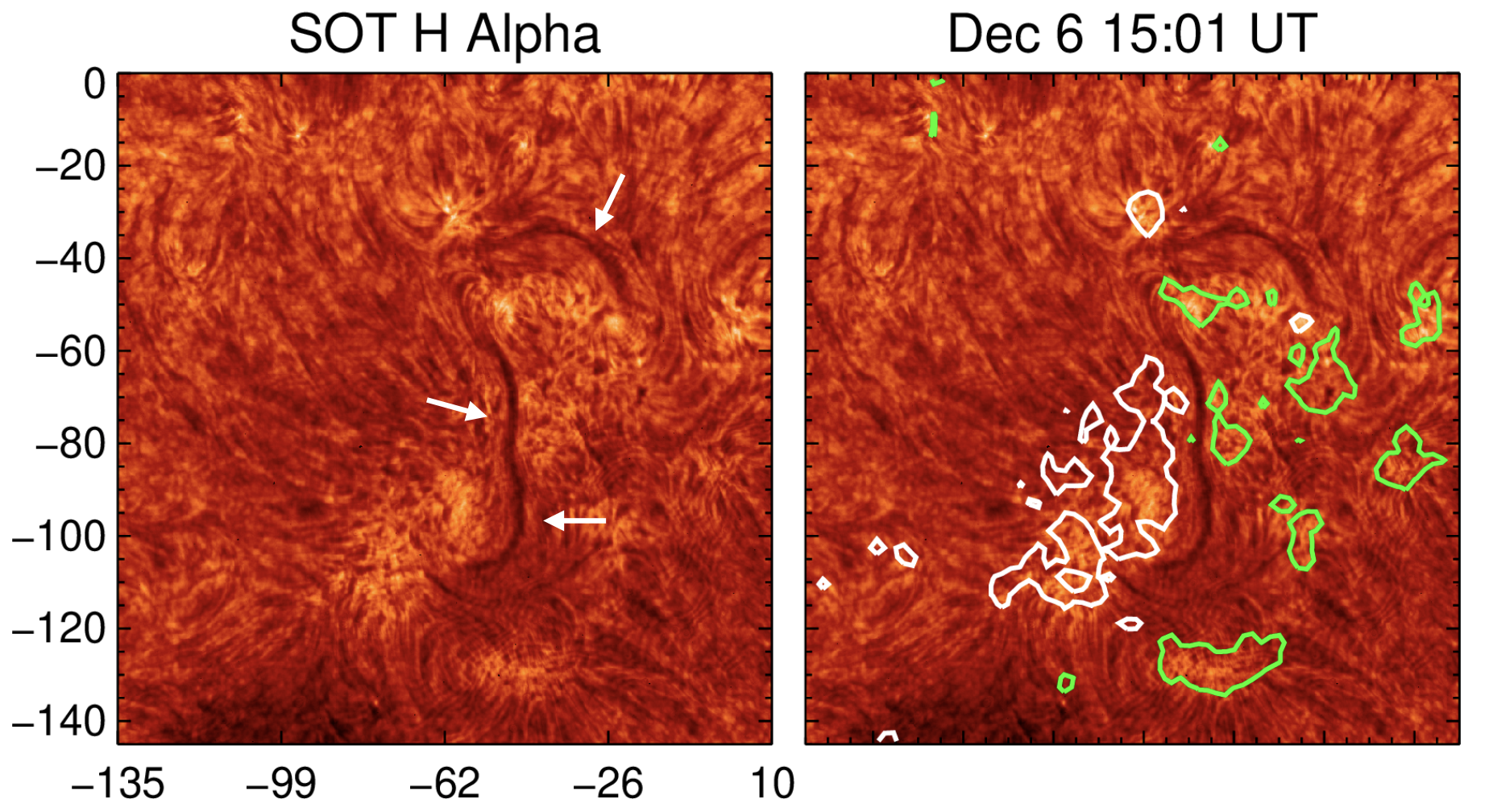}
\caption{ISOON and \emph{Hinode}/SOT H$\alpha$ images at 14:39 UT (a), 17:30 UT (b), and 19:20 UT (c) on December 5 and at 15:01 UT on December 6 without/with contours (left/right) of SOHO/MDI line of sight magnetic field component of $\pm$100 G (white/green for positive/negative values).
Filaments are indicated by the white arrows.
$X$ and $Y$ coordinates are in arcsec.
\label{fig:filament}} 
\end{figure}

A filament is present in AR 10977 on December 5 as shown in the Improved Solar Optical Observing Network (ISOON) and \emph{Hinode}/SOT H$\alpha$ images in Figure \ref{fig:filament}.
It lies along the main PIL and extends to the northwest of the active region.
By the start of the \emph{Hinode}/SOT observing window at 15:01 UT on December 6, a newly formed branch of the filament is observed in the north.
The full extent of the filament then has an S-shape similar to that of the  sigmoid observed in the soft X-ray images in Figure \ref{fig:xrt_series}. 
The distinctive S-shaped filament remains essentially intact during the failed eruption and CME and throughout the 7th (not shown in Figure \ref{fig:filament}).
This is sometimes observed in other events \citep[e.g.][]{dudik14} and it implies that the low lying magnetic configuration supporting the filament is not participating in the eruptions.

\section{\emph{Hinode}/EIS Observations}\label{s:eis}

\emph{Hinode}/EIS observed AR 10977 from 2007 December 5--7, during which time 
Study \#180 was run 16 times, 13 of which are included here.
A field-of-view of 180$\arcsec\times$512$\arcsec$ was constructed by stepping the 1$\arcsec$ slit in 3$\arcsec$ increments for 60 pointing positions, taking 50 second exposures at each position.
All spectra were corrected for dark current,  warm/hot/dusty pixels, and slit tilt using standard EIS routines in the SolarSoft Library \citep{freeland98}.
Single Gaussian functions were fitted to the unblended emission lines \citep{brown08} used for the  composition ratio and temperature measurements.

\begin{figure*}[hbt!]
\epsscale{1.15}
\plotone{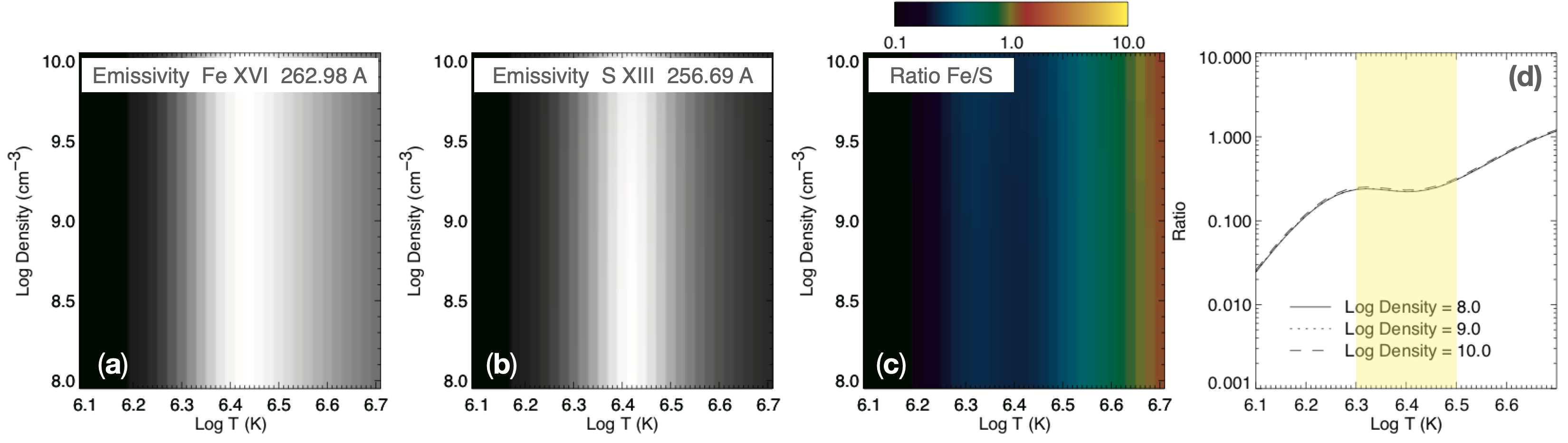}
\caption{\ion{Fe}{16} 262.98 $\angstrom$ / \ion{S}{13} 256.69 $\angstrom$ composition ratio.  Left to right:  Emissivity as a function of temperature and density for \ion{Fe}{16} 262.98  $\angstrom$ (a), \ion{S}{13} 256.69 $\angstrom$ (b), ratio of the Fe/S lines (c), and ratio as a function of temperature for densities of log N$_{e}$ = [8--10] (d).
\label{fig:theory}} 
\end{figure*}

In order to investigate the evolution of coronal plasma composition, a suitable low-FIP and high-FIP spectral line pair must be identified amongst the available lines.
Previous EIS composition studies have employed the \ion{Si}{10} 258.38 $\angstrom$/\ion{S}{10} 264.22 $\angstrom$ line pair for 1--2 MK plasma \citep[e.g.][]{brooks11,baker13,brooks15} and the \ion{Ca}{14} 193.87 $\angstrom$/\ion{Ar}{14} 194.40 $\angstrom$ line pair for 3--4 MK plasmas  \citep[e.g.][]{doschek15,baker20,to21}.
Neither of these well known composition diagnostics is available in Study \#180.
Instead, we use the low-FIP  \ion{Fe}{16} 262.98 $\angstrom$ / high-FIP \ion{S}{13} 256.69 $\angstrom$ line pair recommended by \cite{feldman09} for measuring the FIP effect at temperatures of 2--3 MK, a suitable range for sigmoids \citep[e.g.][]{tripathi09}.
The ratio was computed using the CHIANTI Atomic Database, Version 10 \citep{dere97,gdz21}.
Figure \ref{fig:theory} shows the emissivity of \ion{Fe}{16} 262.98 $\angstrom$ (panel a),  \ion{S}{13} 256.69 $\angstrom$ (panel b), and the Fe/S ratio (panel c) as a function of temperature and density, and the ratio as a function of temperature for specific densities of log N$_{e}$ = [8, 9, 10] (panel d). 
The emissivities were determined using photospheric abundances where log(H) = 12, log(Fe) = 7.45 and log(S) = 7.14 \citep{grevesse07}.

The \ion{Fe}{16} 262.98 $\angstrom$ and \ion{S}{13} 256.69 $\angstrom$ lines have similar temperature dependence in ionization equilibrium \citep{feldman09} and negligible electron density dependence, therefore, their intensity ratio depends primarily on the relative abundances of Fe and S.
In the temperature range log $T_{K}$ = [6.3, 6.5], the relationship is well constrained as shown with the yellow shaded region in panel (d) of Figure \ref{fig:theory}.
In this range, the ratio curves in panel (d) are relatively flat and independent of electron density, with a variation span of 23$\%$.
Therefore, in this study, we use the \ion{Fe}{16} 262.98 $\angstrom$ and \ion{S}{13} 256.69 $\angstrom$ intensity ratio to determine whether the plasma composition is of photospheric or coronal origin and we refer to the spatial distribution of this ratio as composition ratio maps.
A ratio of $\sim$0.20-0.25 is photospheric plasma composition (= unfractionated plasma) and a ratio $>$0.80, a factor over 3$\times$ the photospheric ratio, is coronal plasma composition (= fractionated plasma).
Because of the stronger temperature dependence (see again Figure \ref{fig:theory}(d)), it becomes more difficult to disentangle temperature effects from those of changes in relative abundances for plasma temperatures outside of log $T_{K}$ = [6.3, 6.5].

\begin{figure}[hbt!]
\epsscale{1.1}
\plotone{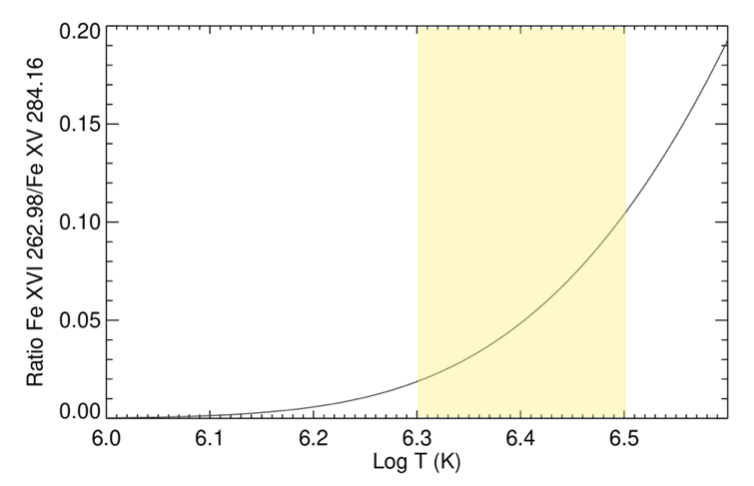}
\caption{Theoretical curve of the \ion{Fe}{16} 262.98  $\angstrom$/ \ion{Fe}{15} 284.16 $\angstrom$ intensity ratio vs. log $T_{K}$.
Temperature range highlighted in yellow is the same as in Figure \ref{fig:theory} (d).   \label{fig:temp}} 
\end{figure}

Temperature maps of the sigmoidal active region were made using the \ion{Fe}{16} 262.98 $\angstrom$ /\ion{Fe}{15} 284.16 $\angstrom$ diagnostic line ratio.
The temperature range covered by the line ratio is compatible with that of the composition diagnostic used in this study.
Figure \ref{fig:temp} shows the theoretical temperature curve derived for the \ion{Fe}{16} 262.98 $\angstrom$ /\ion{Fe}{15} 284.16 $\angstrom$ ratio with the range of log $T_{K}$ = [6.3, 6.5] highlighted in yellow. 
(See the Appendix for \emph{Hinode}/XRT filter ratio temperature maps in support of the \ion{Fe}{16} 262.98 $\angstrom$ /\ion{Fe}{15} 284.16 $\angstrom$ temperature ratio maps.)
The estimated uncertainties in the \ion{Fe}{16} 262.98 $\angstrom$/\ion{S}{13} 256.69 $\angstrom$ composition and the \ion{Fe}{16} 262.98 $\angstrom$/\ion{Fe}{15} 284.16 $\angstrom$ ratios is $\sim$30\% based on the intensity calibration uncertainty of $\sim$23\% \citep{Lang2006}.
 It should be noted that some low intensity pixels may have unreliable
 composition ratio measurements.
Sufficiently low intensities mean that the  \ion{Fe}{16} 262.98 $\angstrom$ and/or \ion{S}{13} 256.69 $\angstrom$ lines may not have well defined spectral profiles over and above the background level.
This is more likely to affect pixels in regions of photospheric composition plasma. 
 
\begin{figure*}[hbt!]
\epsscale{1.1}
\plotone{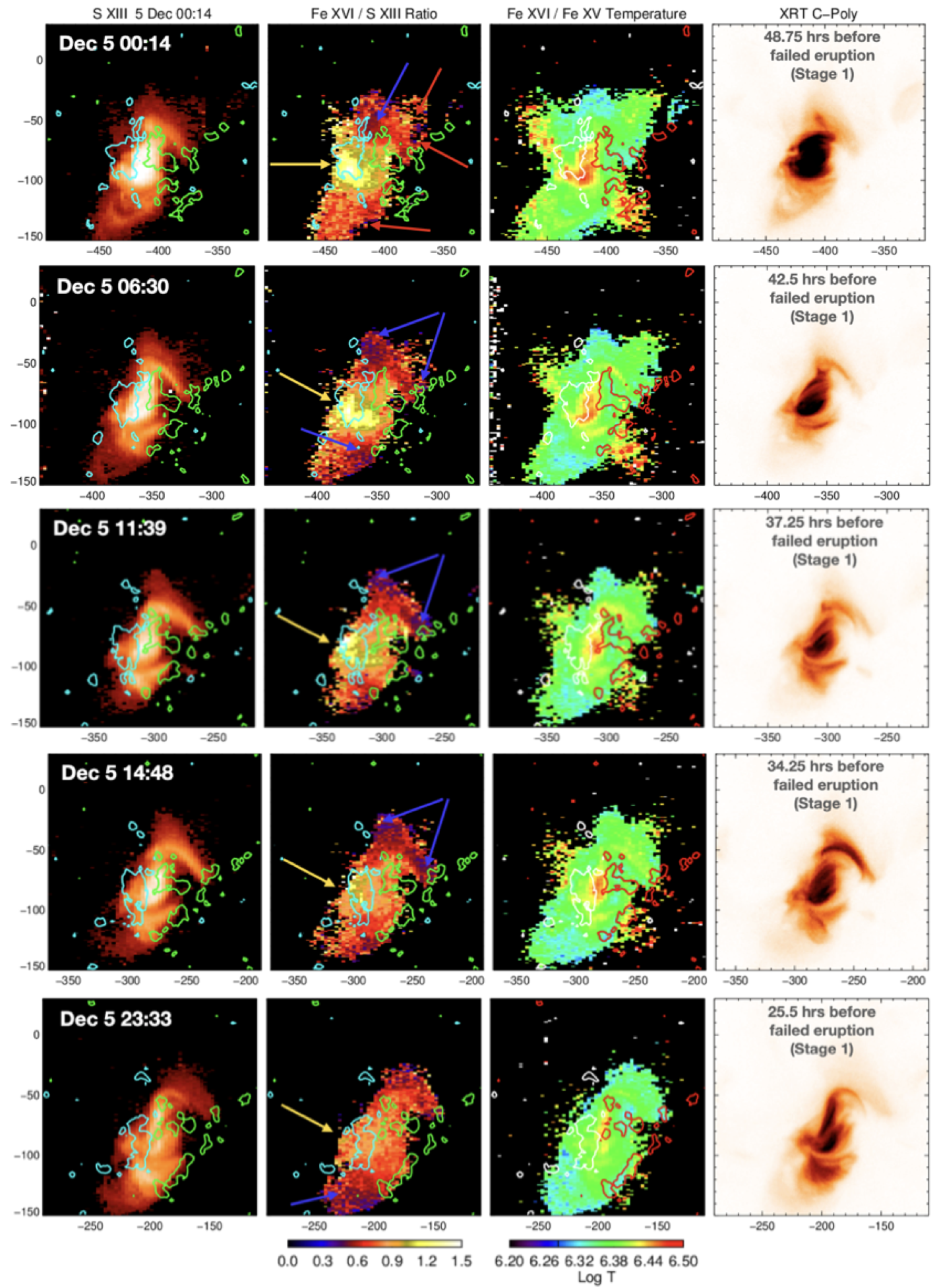}
\caption{\emph{Hinode}/EIS  \ion{S}{13} 256.69 $\angstrom$ intensity, \ion{Fe}{16} 262.98 $\angstrom$/\ion{S}{13} 256.69 $\angstrom$ composition ratio, \ion{Fe}{16} 262.98 $\angstrom$/\ion{Fe}{15} 284.16 $\angstrom$ temperature and XRT intensity maps on 2007 December 5. MDI contours of $\pm$100 G are overplotted on the EIS maps (positive - turquoise or white, negative - green or red). Blue/red/yellow arrows designate unfractionated/partially fractionated/highly fractionated coronal plasmas as discussed in the text. Stages of evolution are explained at the beginning of Section \ref{s:maps}.
$X$ and $Y$ coordinates are in arcsec.
\label{fig:comp1}} 
\end{figure*}

\begin{figure*}[hbt!]
\epsscale{1.2}
\plotone{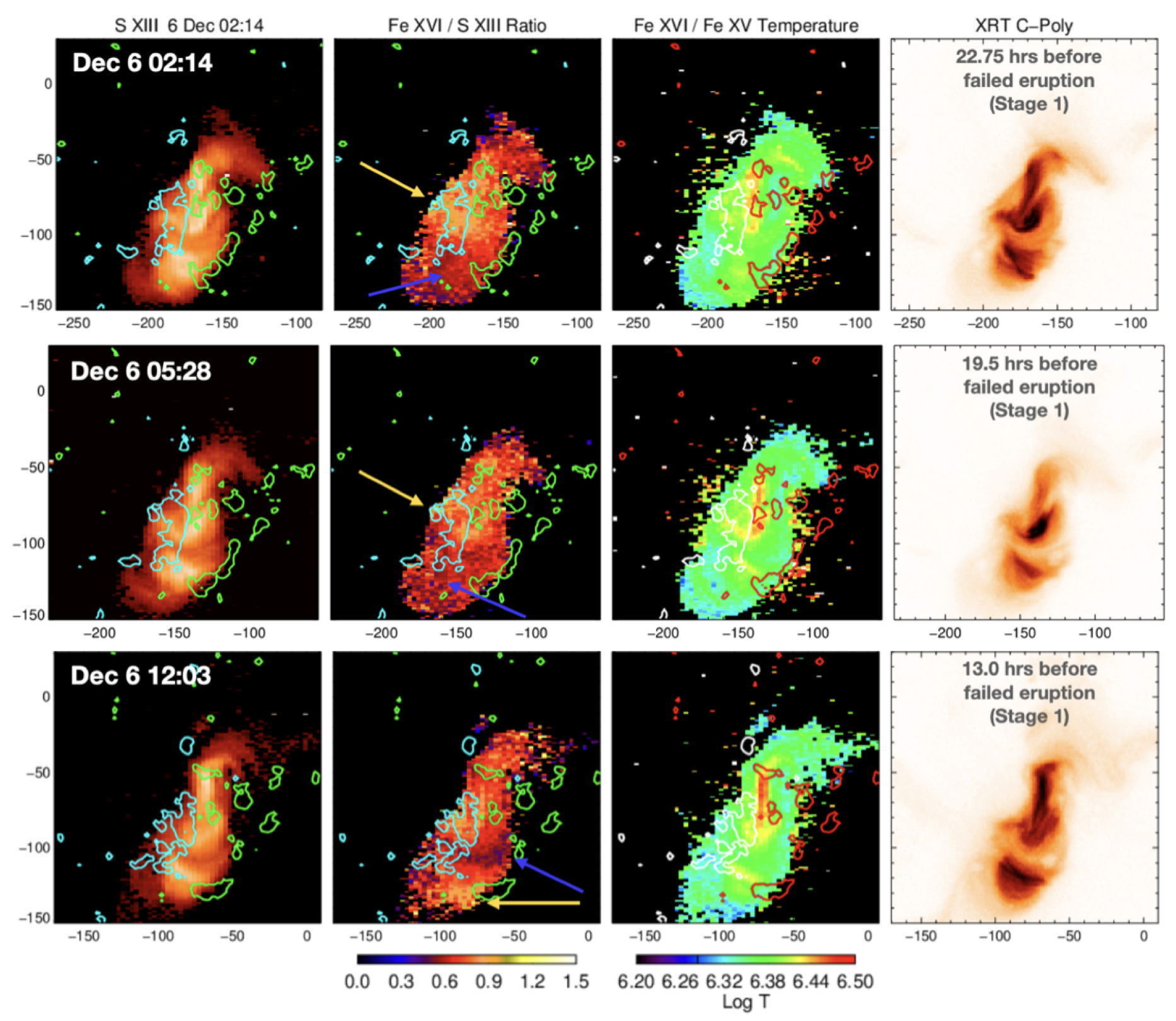}
\caption{\emph{Hinode}/EIS  \ion{S}{13} 256.69 $\angstrom$ intensity, \ion{Fe}{16} 262.98 $\angstrom$/\ion{S}{13} 256.69 $\angstrom$ composition ratio, \ion{Fe}{16} 262.98 $\angstrom$/\ion{Fe}{15} 284.16 $\angstrom$ temperature and XRT intensity maps on 2007 December 6. MDI contours of $\pm$100 G are overplotted on the EIS maps (positive - turquoise or white, negative - green or red). Blue/red/yellow arrows designate unfractionated/partially fractionated/highly fractionated coronal plasmas as discussed in the text.
Stages of evolution are explained at the beginning of Section \ref{s:maps}.
$X$ and $Y$ coordinates are in arcsec.
\label{fig:comp2}} 
\end{figure*}

\begin{figure*}[hbt!]
\epsscale{1.1}
\plotone{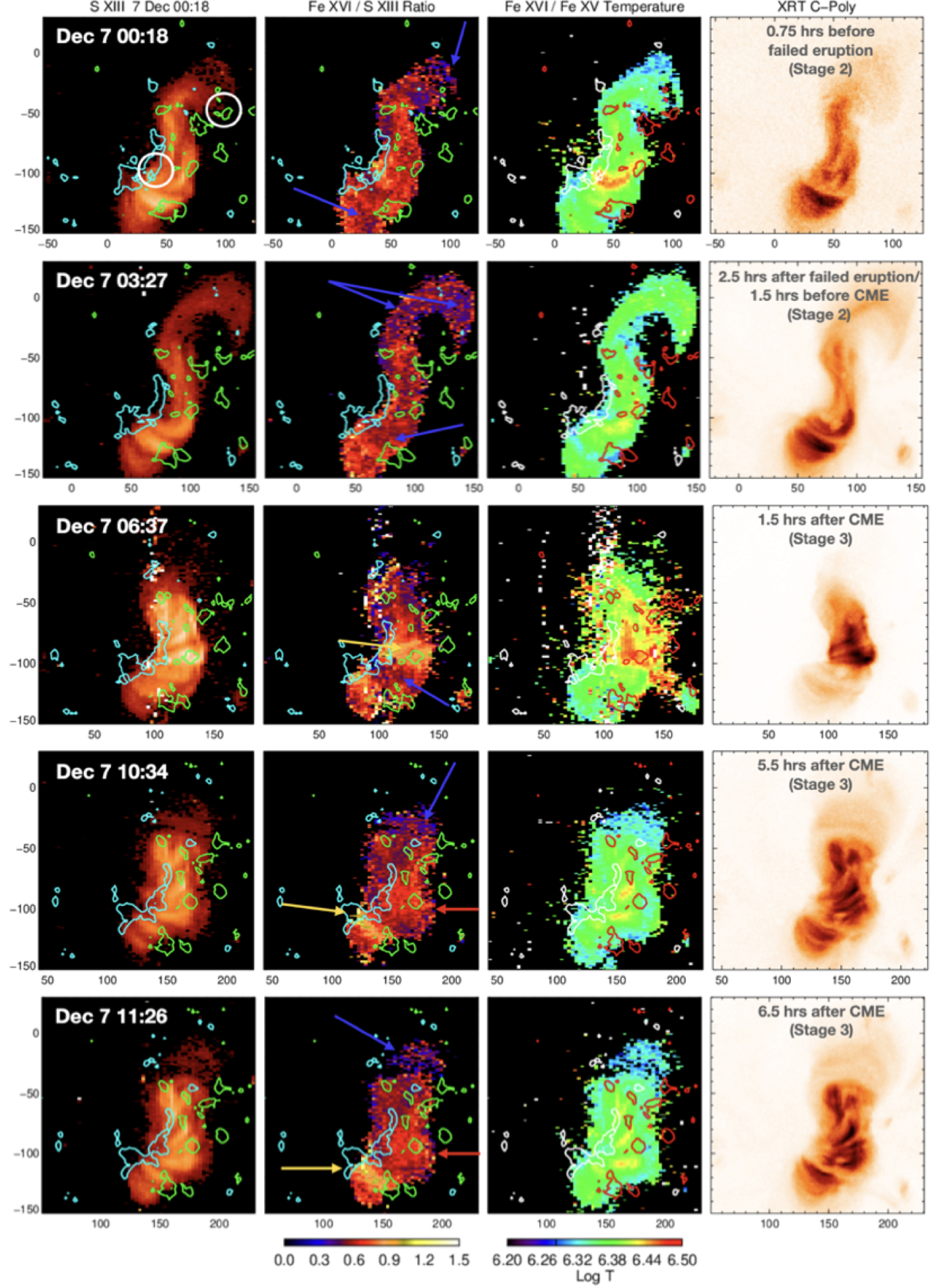}
\caption{\emph{Hinode}/EIS  \ion{S}{13} 256.69 $\angstrom$ intensity, \ion{Fe}{16} 262.98 $\angstrom$/\ion{S}{13} 256.69 $\angstrom$ composition ratio, \ion{Fe}{16} 262.98 $\angstrom$/\ion{Fe}{15} 284.16 $\angstrom$ temperature and XRT intensity maps on 2007 December 7. MDI contours of $\pm$100 G are overplotted on the EIS maps (positive - turquoise or white, negative - green or red). Blue/red/yellow arrows designate unfractionated/partially fractionated/highly fractionated coronal plasmas as discussed in the text.
Stages of evolution are explained at the beginning of Section \ref{s:maps}.
$X$ and $Y$ coordinates are in arcsec.
\label{fig:comp3}} 
\end{figure*}

\emph{Hinode}/EIS \ion{S}{13} 256.69 $\angstrom$ intensity, \ion{Fe}{16} 262.98 $\angstrom$/\ion{S}{13} 256.69 $\angstrom$ composition ratio, and  \ion{Fe}{16} 262.98 $\angstrom$/\ion{Fe}{15} 284.16 $\angstrom$ temperature maps are provided in Figures \ref{fig:comp1}--\ref{fig:comp3}. 
Each map has been overplotted with \emph{SOHO}/MDI $\pm$100 G contours of the line of sight magnetic field component closest in time and differentially rotated to the start time of the EIS raster.
The right column of the figures shows the corresponding \emph{Hinode}/XRT C-poly intensity maps.
The color scheme for the composition ratio maps has been chosen so that photospheric composition with a ratio of $\sim$0.20--0.25 is dark (blue) and coronal composition with a ratio of $>$0.8 is light (tan/yellow/white).
Orange/red indicates mixed or partially fractionated plasma in between photospheric and coronal composition.
The color of the arrows in the composition ratio maps corresponds to this color scheme.

\section{Composition and temperature evolution in the sigmoidal active region}\label{s:maps}
\emph{Hinode}/EIS observations span the three main stages of the coronal evolution in the sigmoidal active region:  Stage 1 - sigmoid/flux rope formation (observations at 00:14, 06:30, 11:39, 14:48, and 23:33 UT on December 5 and 02:14, 05:28, and 12:03 UT on December 6 in Figures \ref{fig:comp1} and \ref{fig:comp2}), Stage 2 - failed eruption and CME (observations at 00:18 and 03:27 on December 7 in Figure \ref{fig:comp3}), and Stage 3 - post-eruptive period (observations at 06:37, 10:34, and 11:26 UT on December 7 in Figure \ref{fig:comp3}).

\subsection{Flux Rope Formation (Stage 1)}\label{stage1}
The northern and southern sections of AR 10977 evolved separately throughout the active region's decay phase.
Bright sheared loops crossed the main PIL,  connecting the main magnetic polarities.  
These loops in the \ion{S}{13} 256.69 $\angstrom$ and XRT intensity images correspond to highly  fractionated plasma in the \ion{Fe}{16} 262.98 $\angstrom$/\ion{S}{13} 256.69 $\angstrom$ composition ratio maps at 00:14, 06:30, and 11:39 UT on December 5, as indicated by the yellow arrows in Figure \ref{fig:comp1}.
The degree of plasma fractionation in these loops notably decreases in the EIS composition ratio maps from 14:48 UT on December 5 and this trend continues on the 6th.

In the region north of the sheared arcade, plasma composition is predominantly partially fractionated (red arrows) at 00:14 UT. 
However, at the localized primary site of flux cancellation (see Figure \ref{fig:mdi_series}), the plasma composition is approaching photospheric levels (blue arrow). 
The spatial distribution of the photospheric-like composition extends away from the primary site of flux cancellation along loops in the northern-most region (called the elbow) in the composition ratio maps at 06:30, 11:38, and 14:48 UT on December 5 (Figure \ref{fig:comp1}).
As the sigmoid develops the northern elbow reverts to partially fractionated (red) plasma on December 6.

The loops in the southern section of the active region remained essentially perpendicular to the PIL, i.e. potential-like, throughout the sigmoid formation.
Plasma composition in this region evolved from partially fractionated (red) at 00:14 UT on December 5 to photospheric-like hours later.
The blue patch is located at the secondary site of flux cancellation.
In general, the southern region has mixed plasma composition until a distinct curved feature of photospheric (blue) plasma develops in the magnetic void between the fragmented polarities at $\sim$--125$\arcsec$ in $Y$ (see the \ion{Fe}{16} 262.98 $\angstrom$/\ion{S}{13} 256.69 $\angstrom$ ratio map at 12:03 UT on December 6 in Figure \ref{fig:comp2}).

The \ion{Fe}{16} 262.98 $\angstrom$/\ion{Fe}{15} 284.16 $\angstrom$ temperature maps in Figures \ref{fig:comp1} and \ref{fig:comp2} show very little temperature evolution within the considered narrow range in the loops to the north and south of the sheared arcade.
Temperatures remain in the range $\log T_K$ = [6.32, 6.38] on December 5 and 6 well inside the reliable temperature window of our diagnostics as marked by the yellow band in Figure \ref{fig:temp}.
Within the sheared arcade, the zone along the main PIL is hotter than its surroundings with temperatures in excess of $\log T_K$ = 6.4, approaching $\log T_K$ = 6.5.
The extent of the hottest region evolves from a broad patch of approximately 30$\arcsec$ $\times$ 40$\arcsec$ to a narrow bar-like feature along the axis of the S-shaped sigmoid as it forms.

\subsection{Eruptions (Stage 2)}\label{stage2}
\emph{Hinode}/EIS observed the expanded and extended sigmoid during the rise phase of the light curve peak associated with the failed eruption at 00:18 UT (top panel of Figure \ref{fig:comp3}).
Overall, plasma composition has evolved from partially fractionated to more photospheric-like at the extreme ends of the S-shaped loops, most notably in the northern elbow.
After the failed eruption but before the CME, the sigmoid is dominated by photospheric composition along its full extent (at 03:27 UT).
The temperature of the flux rope structure is approximately log $T_{K}$ = 6.35 at this time and the hot bar-like feature along the sigmoid's axis is not prominent in the composition ratio or temperature maps, although it is still clear in the \ion{S}{13} 256.69 $\angstrom$ intensity map.

\subsection{Post CME Eruption (Stage 3)}\label{stage3}
The sigmoid structure was destroyed during the CME and replaced by highly sheared loops  \citep[see Figure \ref{fig:comp3} at 06:37 UT; ][]{green11}, while the potential-like arcade in the southern part of the active region remained intact.
The structure of the active region and  distribution of plasma composition are similar to that of the first EIS observation at 00:14 UT on December 5.
Coronal composition plasma comprises the post-eruption arcade loops.
Patches of photospheric composition plasma persist at the secondary site of flux cancellation in the southern half of the active region.
The largest spatial extent of hot plasma (log $T_{K}$ = [6.45, 6.5]) is observed in the post-eruption arcade at 06:37 UT, after which the temperature returns to the characteristic pre-eruption range of log $T_{K}$ = [6.32, 6.38] a few hours later.

\section{Discussion}\label{s:discussion}
In Section \ref{s:maps} we show how plasma composition evolved in an active region that became sigmoidal as its sheared arcade field was transformed into a flux rope that eventually erupted as a CME.
Photospheric composition plasma was found in coronal structures connected to sites of flux cancellation along the main PIL.
Highly fractionated plasma was observed in the sheared arcade field at the beginning of significant flux cancellation on December 5 and then again shortly after the CME on December 7.
Within the temperature range over which the \ion{Fe}{16} 262.98 $\angstrom$/\ion{S}{13} 256.69 $\angstrom$ composition diagnostic is effective (log $T_{K}$ = [6.3, 6.5]), photospheric composition was observed where the plasma temperature was log $T_{K}$ $\lesssim$ 6.35, and coronal composition where the plasma temperature was log $T_{K}$ $\gtrsim$ 6.4, suggesting that the level of fractionation, i.e. coronal plasma composition, is linked to the level of coronal heating.  
More precisely, it is linked to the height where reconnection is occurring (see Sections \ref{s:Photospheric_Cancellation} and \ref{s:Local_Evolution_Sites}). 
When reconnection occurs in the corona, the released energy is transported, e.g. by MHD waves, along the new formed flux tubes. 
This is associated with an increase in the composition ratio (i.e. level of coronal composition plasma). 
If reconnection occurs at the photospheric/chromospheric level, in particular at bald patches, plasma with photospheric composition is injected in the corona.  
This is in agreement with the results of \citet{fletcher01}.

\subsection{Photospheric Flux Cancellation Mechanism of Flux Rope Formation}
\label{s:Photospheric_Cancellation}
The flux rope formation mechanism based on flux cancellation reported in \cite{green11} and modelled in \cite{aulanier10}, \cite{savcheva12} and \cite{gibb14} provides the basis for our understanding of the evolution of plasma composition and temperature in AR 10977.
\cite{martin85} define flux cancellation as the apparent loss of magnetic flux in closely spaced magnetic field of opposite polarities.  
Photospheric converging motions towards the PIL is a natural process occurring as a consequence of the active region's magnetic field dispersion driven by convection \citep[][and references therein]{lvdg15}.
Flux cancellation in a sheared arcade is the observational manifestation of reconnection along the PIL which in turn alters the coronal magnetic field structure.

A description of the process from \cite{vanballegooijen89} is summarized below.
The process is illustrated in their Figure 1.
As two sheared loops crossing the PIL have one of their footpoints, of opposite magnetic polarity, moving towards each other by the converging motions, they are forced to reconnect. 
This process creates two new loops - a short loop that submerges at the PIL and a long loop connecting the distant footpoints.
The submerged field reduces the flux content within the system which is detected as flux cancellation. 
 The converging motions also increase the magnetic shear of the loops before reconnection, as well as the shear of the long loops formed by reconnection. 
 It results in an increasingly sheared magnetic arcade, with a more sheared core, and later on a flux rope is formed.  
 This later stage is due to less sheared loops reconnecting with more sheared loops below them, so that the reconnected loops are forced to wrap around, with a helix-like shape, the more sheared loops (which become the core of the flux rope).  
 This process, envisioned first by \cite{vanballegooijen89} is presently confirmed by numerical simulations \citep[e.g.][]{amari03, aulanier10, amari11, zuccarello15}. 

The overlying arcade loops keep the flux rope line-tied at the photosphere forming a bald-patch separatrix surface \citep[BPSS;][]{titov93}.
Where the helical field grazes the photosphere at the BPSS, the field lines are concave up, forming magnetic dips in which dense filament material can be supported along the PIL.
Dips occur at the center of the S-shaped field located at sites of flux cancellation \citep{titov99, savcheva12}.
The flux cancellation process adds flux to the flux rope and, as the flux rope is forming, it is moving upward to satisfy the force balance. 
As a consequence of the partial detachment of the flux rope from the photosphere, the BP first splits into two BPs which progressively separate with time.  
The two associated BPSS intersect at a coronal separator where magnetic reconnection is also forced to occur. 
Later on,  as the flux rope is rising, BPs disappear and then magnetic dips are  present only at the coronal level. 
The separator / intersecting BPSS is transformed to a hyperbolic flux tube (HFT) / QSLs \citep[e.g. Figure 4 of][]{aulanier10}.  

Therefore, during this entire process, converging photospheric motions with flux cancellation first impose reconnection of field lines at the photospheric level, then at the coronal level. 
This creates the envelope of the flux rope, further building it. 
Eventually, the overlying arcade can no longer hold down the flux rope due to  built-up magnetic pressure and it subsequently erupts.

\subsection{Local Evolution - Sites of Flux Cancellation}
\label{s:Local_Evolution_Sites}
AR 10977 exhibited significant photospheric flux cancellation along the internal or main PIL beginning early on 2007 December 5 (see Figure \ref{fig:global}(a)).
The primary site of flux cancellation is in the northern region where the flux rope formed \citep{green11,savcheva12}.
Photospheric plasma composition is first observed at the precise location of cancellation at the time when the flux curve in Figure \ref{fig:global}(a) is steepest, suggesting a fast rate of flux cancellation along the PIL.
Composition ratio maps timed at 00:14, 06:30, and 11:39 UT coincide with the sharp fall in flux from approximately 00:00 to 15:00 UT on December 5.
In these maps, the area in the corona containing plasma with photospheric composition, extends further north-northwest where the loops of the northern elbow are located.
Over the same time period, a number of flaring events are observed in the enclosed XRT movie called `XRT$\_$movie.mp4'.
These events are likely to be related to reconnection induced by the ongoing flux cancellation and subsequent reorganization of the coronal field in the northern region.
Loops containing photospheric composition rooted in and around the primary site of flux cancellation are able to reconnect with nearby loops thereby transferring the plasma over a larger area as observed by \emph{Hinode}/EIS.
By the same mechanisms the plasma with photospheric composition contained in the flux rope is heated up by small-scale reconnection events so that it appears in the temperature window observed by EIS.

Flux cancellation at the secondary site (see Figure \ref{fig:mdi_series}) is weaker and begins somewhat later compared to that of the primary site \citep{green11}.
However, a similar evolution of plasma composition is observed in the far south of the active region.
Figure \ref{fig:comp1} at 23:33 UT shows a region of photospheric plasma composition at the secondary flux cancellation site.
Eruptive activity is observed in the southern region in the XRT movie during December 5 leading up to the EIS observation.
The scenario is comparable to that in the northern section of the active region.

Unfractionated plasma composition at the sites of flux cancellation is consistent with a BPSS topology where the field lines are tangential to the photosphere.
The simulations of AR 10977 carried out by \cite{gibb14} support the presence of a BPSS topology as the flux rope is very low down, forming at a height of 2 Mm.
BPSS are locations where current sheets form and reconnection takes place, albeit very low in the solar atmosphere.
During reconnection at the BPSS, the energy released along the field lines causes heating and evaporation of photospheric plasma.
After reconnection at the BPSS, photospheric plasma can be lifted into the corona as the concave up BPSS field lines rise \citep{titov93,fletcher01,aulanier10}.
This scenario of AR 10977 is similar to that reported in  \cite{baker13} where
\emph{Hinode}/EIS observed photospheric plasma composition  along the sigmoidal channel above the main PIL hours before a CME.
The results of \cite{fletcher01} confirm the link between photospheric plasma composition, locations of flux cancellation, and a BPSS topology.
They found that the elemental abundances measured in the transition region brightenings within a sigmoidal active region depends on the type of topological structure of the regions where the brightenings occur; coronal plasma with photospheric composition was associated with BPSS and coronal plasma with quasi-separatrix layers (QSLs). 

\cite{savcheva12} (Figure 11) modelled the sigmoid/flux rope just prior to the composition ratio map at 12:03 UT and identified the locations of flux rope associated field line dips within AR 10977.
One dip is located at the primary site of flux cancellation and the other is in the magnetic void, i.e. the region of low radiance in the band of photospheric composition that is indicated by the blue arrow at 12:03 on December 6 (Figure \ref{fig:comp2}). 
It is tempting to claim that the photospheric plasma composition is directly related to the magnetic dip at the BPSS identified by \cite{savcheva12}.
However, it is equally plausible that the unfractionated plasma is also related to the low amount of coronal heating, as traced by the low temperature and density present there, leading to a lower amount of MHD waves and a low level of the fractionation process of the model developed by \citet{laming15}.  

Finally, early studies \cite[e.g.][]{spicer98} found that plasma composition in prominences (and therefore filaments) is photospheric.
On December 5, photospheric composition was observed intermittently along the northern pathway of what would become the S-shaped filament observed on December 6 (Figures \ref{fig:comp1} and \ref{fig:comp2}),
suggesting the possibility that \emph{Hinode}/EIS was observing the plasma of the filament cavity before filament formation in the northern section of the active region.

\subsection{Local Evolution - Arcade Field}
\label{s:Local_Evolution_Arcade}
Within the central arcade field of the active region, highly fractionated plasma with coronal composition is observed during the period of accelerated flux cancellation early on December 5.  
The \ion{Fe}{16} 262.98 $\angstrom$/\ion{S}{13} 256.69 $\angstrom$ ratio evolves to lower levels, though remaining coronal in composition, as the flux cancellation curve flattens at $\sim$15:00 UT.
As the composition ratio decreases, the temperature of the plasma within the arcade field region also decreases.
Concurrent evolution in composition and temperature suggests that reconnection induced by flux cancellation is also decreasing.
Enhanced levels in the low-FIP \ion{Fe}{16} relative to the high-FIP \ion{S}{13} in the arcade field connecting opposite polarities are predicted by the ponderomotive fractionation model of \cite{laming15} and supported by the simulations of \cite{dahlburg16}.
A high Alfv\'en wave flux is expected while magnetic field flux cancellation induced coronal reconnection is ongoing. 
This is in contrast with the description of the scenario in Section \ref{s:Local_Evolution_Sites} where reconnection occurs at the photospheric level. Later on, the Alfv\'en wave flux is expected to fall off with the lower levels of induced reconnection starting later on December 5 and continuing until after the eruptions.
Plasma mixing is likely to occur as arcade loops reconnect with loops rooted in the vicinity of flux cancellation regions thereby contributing to the decrease in \ion{Fe}{16} 262.98 $\angstrom$/\ion{S}{13} 256.69 $\angstrom$ ratio values.
In a similar fashion, coronal composition is observed in the bright, hot post-eruption loops at 06:37 UT on December 7, thereafter, the flare arcade fades, the temperature decreases, and the plasma composition becomes mixed, i.e. partially fractionated (red).

\subsection{Local Evolution - Flux Rope}
\label{s:Local_Evolution_Sigmoidal}
\cite{green11} reported that the flux rope, as traced by coronal plasma, had formed approximately eight hours before the start of the failed eruption.
The footpoints of the flux rope are identified by the white circles overplotted on the \emph{Hinode}/EIS map at 00:18 UT on December 7 (Figure \ref{fig:comp3}).
At this stage, the plasma composition at  each footpoint is predominantly photospheric while partially fractionated plasma (red) comprises the central portion of the flux rope.
Three hours later, after the failed eruption but before the CME, photospheric composition has spread along the entire length of the flux rope (at 03:27 UT in Figure \ref{fig:comp3}).
Prior to and during the eruptive period the sigmoid/flux rope expands and rises  \citep[see Figure 9(a) of][]{gibb14}, allowing heated plasma with photospheric composition at the footpoints to expand into the increasing volume. 
Well before the eruption, reconnection at bald patches is also expected to bring new plasma with photospheric composition at the periphery of the forming flux rope.
As the erupting system enters a phase of fast expansion \citep{aulanier10}, the photospheric plasma is accelerated into the flux rope so that more of the volume is  filled, similar to what is observed at 03:27 UT.
The dominantly photospheric plasma composition of the erupting sigmoid/flux rope is in agreement with the photospheric origin material in erupting  \citep[e.g.][]{widing86,feldman92,spicer98} and quiescent prominences/filaments \citep{parenti19}.

\subsection{Global Evolution}
\label{s:Global_Evolution}
Locally, the spatial distribution and temporal evolution of plasma composition observed in AR 10977 are in support of the scenario proposed by \cite{vanballegooijen89} that flux cancellation at the main PIL of a sheared arcade field leads to the formation of a flux rope.
The global evolution of plasma composition within the active region also appears to be dominated by the processes of flux rope formation in flux cancellation models.
In Figure \ref{fig:global}(a), the mean active region \ion{Fe}{16} 262.98 $\angstrom$/\ion{S}{13} 256.69 $\angstrom$ ratios within the temperature range log $T_{K}$ = [6.3, 6.5] are compared to the flux and temperature evolution.
The mean value of the composition ratio decreases $\sim$30$\%$ during the flux rope formation and eruptions until a few hours after the CME when the active region is dominated by hot, post-eruption loops.
Over the same time period, the magnetic flux decreased by a similar amount, $\sim$29$\%$.

The parallel evolution of plasma composition and magnetic flux is consistent with the conclusions of \cite{baker18}.
They found that the relative abundance of low-FIP \ion{Si}{10} 258.38 $\angstrom$ compared to high-FIP \ion{S}{10} 264.22 $\angstrom$ increases during the magnetic emergence phase and decreases during the magnetic decay phase for active regions ranging from ephemeral flux regions to the largest active regions. 
The strong correlation of composition with magnetic activity extends to solar-cycle time scales \citep{brooks17}.
The mean temperature of the active region remained stable during the flux rope formation followed by the eruptive period, in agreement with the results of \cite{uu12} who demonstrated that active region cores become fainter and less variable during the decay phase.

\section{Conclusion}\label{s:conclusion}
Locally and globally, the \emph{Hinode}/EIS plasma composition and temperature observations of AR 10977 strongly support the
\cite{vanballegooijen89} model of flux rope formation by photospheric flux cancellation.
We employed a new composition diagnostic, the \ion{Fe}{16} 262.98 $\angstrom$/\ion{S}{13} 256.69 $\angstrom$ ratio proposed by \cite{feldman09} to investigate the formation and evolution of a flux rope in a sigmoidal active region.
Our results demonstrate that plasma composition provides independent observational evidence to distinguish between the mechanisms of flux rope formation - those that form via reconnection in the corona \citep[e.g.][]{james17,james18} and those that form lower in the atmosphere via photospheric flux cancellation \citep[e.g.][]{vanballegooijen89,aulanier10}.

\begin{figure*}[hbt!]
\epsscale{1.2}
\plotone{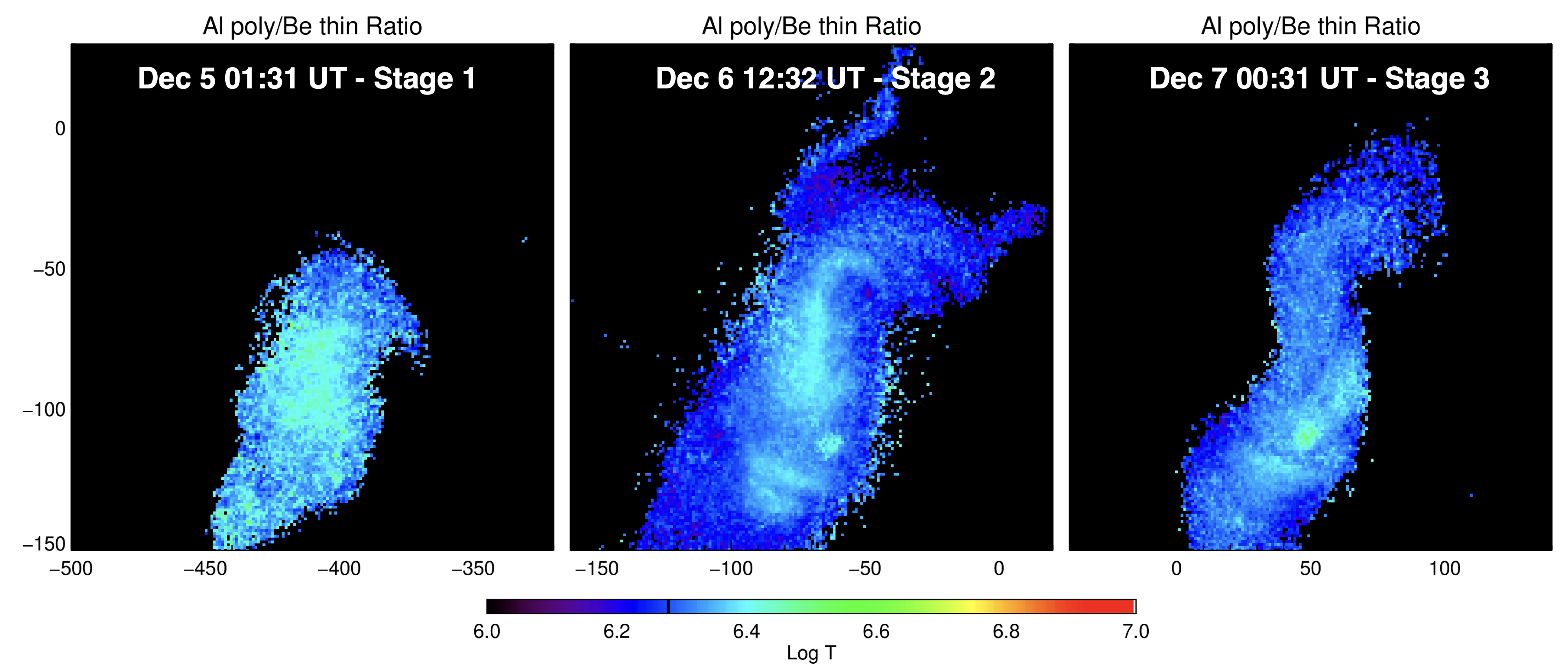}
\caption{\emph{Hinode}/XRT Al poly/Be Thin temperature ratio maps during the stages of evolution of the sigmoid/flux rope.  In all cases the temperature peaks at approximately log $T$ = 6.45.
\label{fig:append}} 
\end{figure*}

\appendix
\cite{mulay21} analyzed the temperature evolution of the hot plasma in a flaring active region containing a sigmoid.
Their study used an emission measure analysis and various filter ratio methods to obtain the temperature distribution in the active region.
They found good agreement between the results using the different methods.
We have used the \emph{Hinode}/XRT Al poly/Be thin filter ratio employed by \cite{mulay21} to compare with the \ion{Fe}{16} 262.98 $\angstrom$/\ion{Fe}{15} 284.16 $\angstrom$ maps in Figures \ref{fig:comp1}--\ref{fig:comp3}.
Figure \ref{fig:append} shows three filter ratio maps, one from each evolutionary stage described in Section \ref{s:maps}.
The temperature distribution in the active region is in the range log $T$ = [6.30, 6.45], consistent with the \ion{Fe}{16}/\ion{Fe}{15} maps.
These results provide a good level of confidence that the \ion{Fe}{16}/\ion{S}{13} composition ratio is a suitable diagnostic for analyzing plasma composition in sigmoidal active regions.

\acknowledgments
We thank the reviewer, Jaroslav Dudik, for his very constructive report that helped to improve the manuscript.
Hinode is a Japanese mission developed and launched by ISAS/JAXA, collaborating with NAOJ as a domestic partner, and NASA and STFC (UK) as international partners. 
Scientific operation of Hinode is performed by the Hinode science team organized at ISAS/JAXA. 
This team mainly consists of scientists from institutes in the partner countries. 
Support for the post-launch operation is provided by JAXA and NAOJ (Japan), STFC (UK), NASA, ESA, and NSC (Norway). 
D.B. is funded under STFC consolidated grant number ST/S000240/1 and L.v.D.G. is partially funded under the same grant.
The work of D.H.B. was performed under contract to the Naval Research Laboratory and was funded by the NASA Hinode program. 
L.v.D.G. acknowledges the Hungarian National Research, Development and Innovation Office grant OTKA K-131508.
S.L.Y. would like to thank NERC for funding via the SWIMMR Aviation Risk Modelling (SWARM) Project, grant number NE/V002899/1. 
D.M.L. is grateful to the Science Technology and Facilities Council for the award of an Ernest Rutherford Fellowship (ST/R003246/1).
We recognise the collaborative and open nature of knowledge creation and dissemination, under the control of the academic community as expressed by Camille No\^{u}s at http://www.cogitamus.fr/indexen.html.

\bibliographystyle{aasjournal}
\bibliography{references}

\end{document}